\documentclass[iop]{emulateapj}
\usepackage{apjfonts}
\usepackage{graphics}
\usepackage{natbib}
\usepackage{amsmath}
\usepackage[version=3]{mhchem}
\newcommand{\fug}{$\mathit{f}$O$_{2}$ }
\newcommand{\fugc}{$\mathit{f}$O$_{2}$}
\newcommand{\sid}{FeCO$_{3}$ }
\newcommand{\sidc}{FeCO$_{3}$}
\slugcomment{Submitted to ApJ October 30, 2013}

\begin{document}

\title{The Role of Carbon in Extrasolar Planetary Geodynamics and Habitability}
\author{Cayman T. Unterborn\altaffilmark{1,\textasteriskcentered}, Jason E. Kabbes\altaffilmark{1}, Jeffrey S. Pigott\altaffilmark{1}, Daniel M. Reaman\altaffilmark{2} and Wendy R. Panero\altaffilmark{1}}
\altaffiltext{1}{School of Earth Sciences, 
The Ohio State University, 125 South Oval \indent \indent Mall, Columbus, OH 43202 USA}
\altaffiltext{2}{US Army Research Laboratory, RDRL-WML-B (Bldg 390), Aberdeen \indent \indent Proving Ground, MD 21005}
\altaffiltext{\textasteriskcentered}{contact: cayman.unterborn@gmail.com}

\begin{abstract}

The proportions of oxygen, carbon and major rock-forming elements (e.g. Mg, Fe, Si) determine a planet's dominant mineralogy. Variation in a planet's mineralogy subsequently affects planetary mantle dynamics as well as any deep water or carbon cycle. Through thermodynamic models and high pressure diamond anvil cell experiments, we demonstrate the oxidation potential of C is above that of Fe at all pressures and temperatures indicative of 0.1 - 2 Earth-mass planets. This means that for a planet with (Mg+2Si+Fe+2C)/O > 1, excess C in the mantle will be in the form of diamond. We model the general dynamic state of planets as a function of interior temperature, carbon composition, and size, showing that above a critical threshold of $\sim$3 atom\% C, limited to no mantle convection will be present assuming an Earth-like geotherm. We assert then that in the C-(Mg+2Si+Fe)-O system, only a very small compositional range produce habitable planets. Planets outside of this habitable range will be dynamically sluggish or stagnant, thus having limited carbon or water cycles leading to surface conditions inhospitable to life as we know it.

\end{abstract}

\section{Introduction}
\label{sec:intro}
The first planet beyond our own solar system was discovered in 1992 \citep{Wols92}. In the nearly two decades since this discovery, over 750 extrasolar planets have been confirmed, with technological breakthroughs allowing for the discovery of planets with masses between that of Mars and Neptune. While most planets orbit stars similar to our own, there is diversity in stellar compositions, particularly in Mg/Si and C/O ratios \citep{Mena10, Bond12a,Fort12, John12,Niss13,Hern13}.  With variable stellar compositions, it is likely that the planets forming from these systems will be similarly variable in bulk composition. Bulk compositional variance in planets will lead to diversity in mantle and core compositions of terrestrial planets \citep{ElkT08}. Furthermore, N-body planetary formation models \citep{Bond10, Bond12b} propose a wide variety of possible bulk planetary compositions as a function of stellar composition.   \\
\indent The interior dynamics of a planet, including the presence of plate tectonics, volcanism, deep volatile cycles and the presence of a magnetic field, are dependent upon how heat is transported from a planetary interior to its surface. For a given planetary radius, the potential for interior mantle convection is quantified by the Rayleigh number \citep{Schu79,Schu80}. The Earth's mantle is a vigorously convecting fluid over its lifetime. In contrast, however, a mantle with a significantly greater viscosity, coupled with high thermal conductivity, creates a system whereby heat is more efficiently transported via radiation or conduction rather than by interior convection. As viscosity and thermal conductivity are physical properties that are a function of both composition and mineral structure, the dynamics of terrestrial planets will then be a function not only of the bulk composition, but the mineral host of the major elements.\\
\indent The mineral hosts of the major elements in a planetary interior are determined by the pressure and temperature-stability field of the structures themselves and the oxidation state of the system. Oxygen fugacity (\fugc) quantifies the oxidation potential of a system, controlling the oxidation state of the elements in the system and therefore the dominant mineral assemblage present in planetary interiors. \\
\indent Carbon exists as diamond throughout much of the Earth's mantle, oxidizing to carbonate at near surface pressures \citep{Dasg10}.  The viscosity and thermal conductivity of diamond, while large compared to carbonates \citep{Clau95}, has very little effect on the dynamics of the Earth because it is present in such small quantities (Table \ref{tab:compcompare}). However, if a planet contains a considerable amount of carbon relative to the other elements present, the role of carbonate or diamond in planetary dynamics cannot be ignored. \\
\indent Static compression and shock wave experiments \citep{Biel93,Issh04,Seki06} suggest that carbonate species are stable over the pressure and temperature range of the Earth's mantle. Oxidized carbon (C$^{4+}$) is not readily incorporated in mantle silicates \citep{Kepp03, Pane08, Tao13}, and will therefore be present as carbonate in planetary mantles if the environment is sufficiently oxidized. Under sufficiently reducing conditions, however, carbon will be in the form of either diamond or carbide \citep{Dasg10,Holl98,Walt08} where the mineralogy is determined by the relative stability of each phase in regards to temperature ($\mathit{T}$), pressure ($\mathit{P}$), and \fugc. Therefore, the oxidation state of carbon within a planetary mantle must be determined before discussing the dynamics of such planets. \\
\indent Planets with high C abundances relative to Earth have been discussed as a tool to explain the mass and radius of one observed planet \citep{Madh12}. We focus, then, on the more general case of the effects of varying planetary carbon content on the dynamics of the planet and its effect on the potential for geologic habitability. In section \ref{sec:OxState} we discuss a model for planetary oxidation in the Fe-C system with section \ref{sec:Exp} providing experimental support to the model. Section \ref{sec:model} describes the implications of an increased concentration of C with respect to the Rayleigh number.\\
\begin{deluxetable}{lc|c}
\tablecolumns{3}
\tabletypesize{\scriptsize}
\tablewidth{0pt}
\tablecaption{Composition of a modeled planet orbiting HD19994 \citep{Bond10} with C/O = 0.78. Bulk Earth values are listed as reference \citep{McD95}}
\tablehead{\colhead{}&\colhead{Earth}&\colhead{HD19994}\nl \colhead{Element}&\colhead{atom\%}&\colhead{atom\%}}
\startdata
Fe &15.7&9.47\nl
Mg &17.4&11.4\nl
Si &15.7&11.1\nl
C &0.17&29.9\nl
O &51.0&38.1 
\enddata
\label{tab:compcompare} 
\end{deluxetable}
\begin{deluxetable}{lc}
\tablecolumns{2}
\tablecaption{Reactions considered in the fugacity model}
\tablehead{\colhead{Name} & \colhead{Reaction}}
\startdata
IW & Fe + 1/2O$_{2}$ $\leftrightarrow$ FeO \nl
CCO (DWS) & FeO + C + 3/2O$_{2}$ $\leftrightarrow$ FeCO$_{3}$ \nl 
MP & Mg + 1/2O$_{2}$ $\leftrightarrow$ MgO \nl
DPM &  MgO + C + 3/2O$_{2}$ $\leftrightarrow$ MgCO$_{3}$ \nl
FMQ & 3Fe$_{2}$SiO$_{4}$ + O$_{2}$ $\leftrightarrow$ 2Fe$_{3}$O$_{4}$ + 3SiO$_{2}$ (stishovite) \nl 
EMOD & Mg$_{2}$SiO$_{4}$ + O$_{2}$ + C $\leftrightarrow$ MgSiO$_{3}$ + MgCO$_{3}$  \nl
EMOD ($>$25 GPa) & MgSiO$_{3}$(pv) + MgO + O$_{2}$ + C  $\leftrightarrow$\nl
&  MgCO$_{3}$ + MgSiO$_{3}$ (pv)
\enddata
\label{tab:reac} 
\end{deluxetable} 

\section{Oxidation State of Carbon Planets}
\label{sec:OxState}
\begin{figure}
\includegraphics[width=8cm]{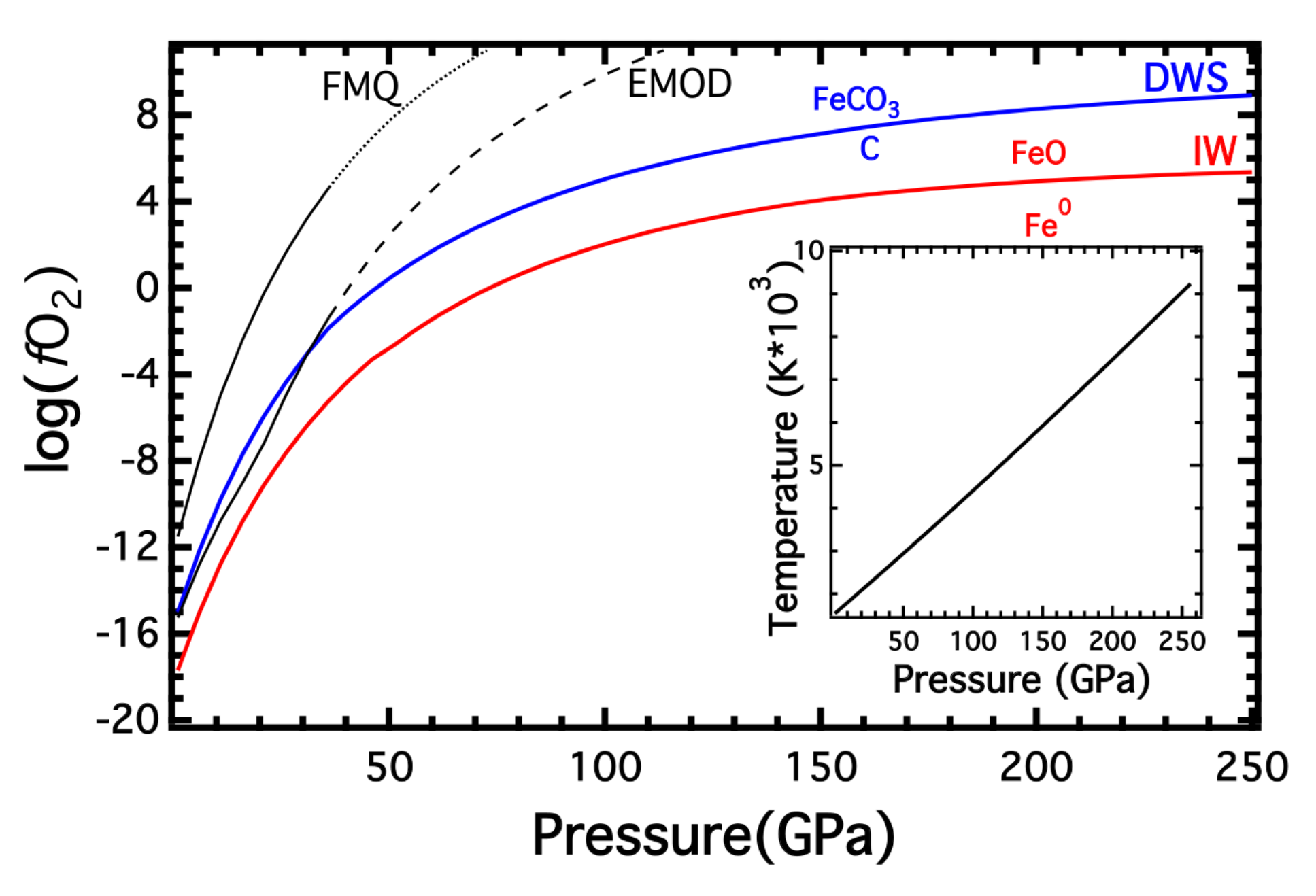}
\caption{Iron to w\"ustite (IW, red) and diamond and w\"ustite to siderite (DWS, blue) fugacities along an adiabatic geotherm (inset) characteristic of an Earth-like silicate mantle. Extrapolated fayalite to magnetite and quartz (FMQ, black dotted) and olivine and diamond to enstatite and magnesite (EMOD, black dashed) fugacities are shown for reference with stable mineral assemblages shown as thin solid black lines.}
\label{fig:Fug_final}
\end{figure}

\begin{figure}
\includegraphics[width=9cm]{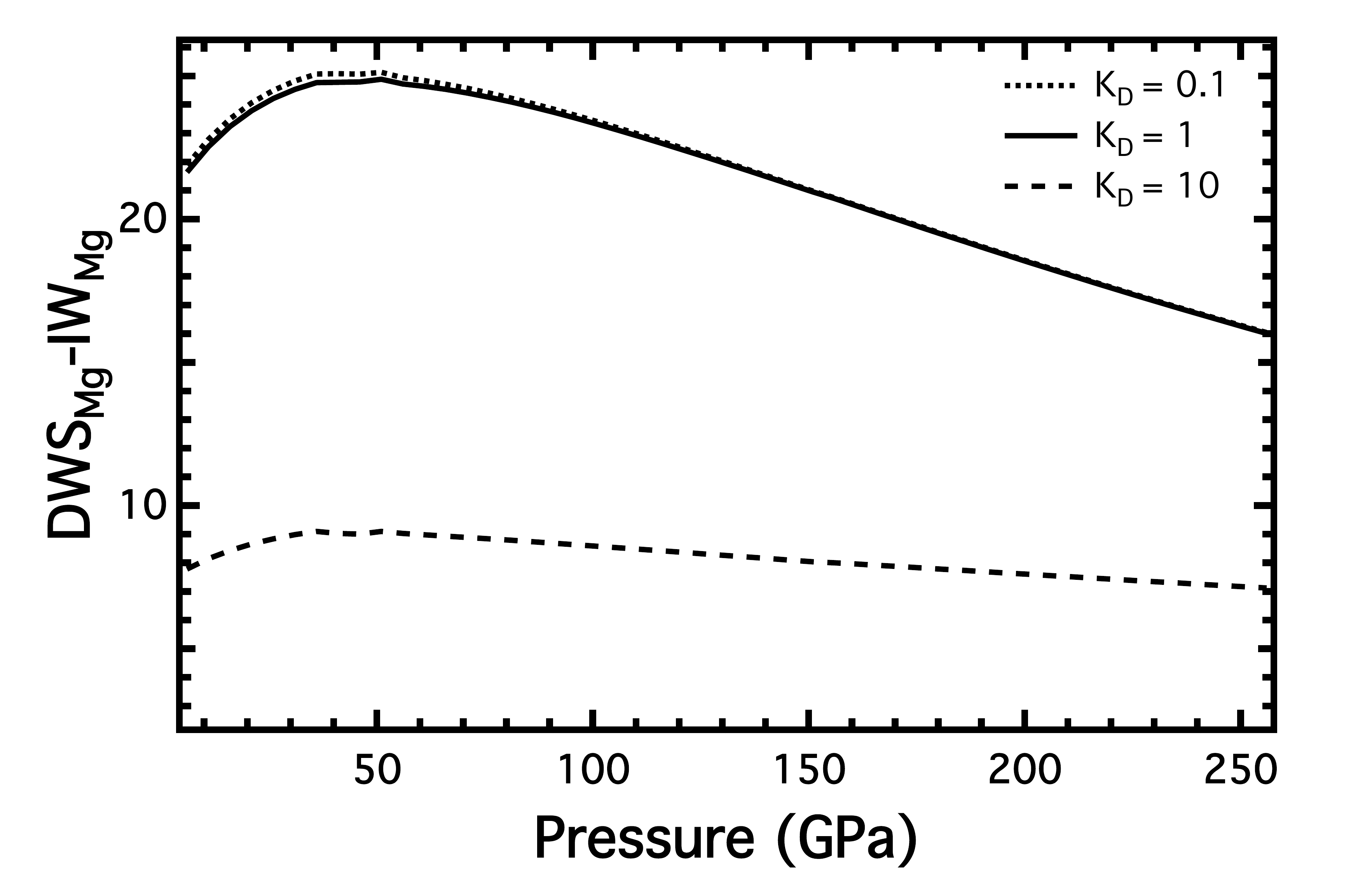}
\caption{The difference between the DWS (CCO) and IW fugacities including the effects of Mg$^{2+}$ substitution for Fe$^{2+}$ at K$_{D}^{Mg}$ = 1 (solid), K$_{D}^{Mg}$ = 0.1 (dotted) and K$_{D}^{Mg}$ = 10 (dashed). }
\label{fig:mixing}
\end{figure}

\indent The mineral host of carbon is either oxidized carbonates (e.g. \sidc, siderite) or the reduced native element (graphite, diamond, alloys) and is a function of the planetary mantle's \fugc. The relative \fug of the carbon-carbonate (CCO) and iron-w\"{u}stite (IW) oxidation potentials determines the stabilities of the C and Fe bearing minerals at the relevant pressures and temperatures. \\ 
\indent We model CCO oxidation potential as the oxidation of diamond and w\"{u}stite (FeO) to siderite (\sidc) reaction as a function of $P$ and $T$ (DWS). The \fug at equilibrium at pressure ($\mathit{P}$) and temperature ($\mathit{T}$) for the reactions in Table \ref{tab:reac} is
\begin{equation}
log(fO_{2})=y\left ( \frac{\Delta H +\int_{P_{0}}^{P}\Delta V dP}{2.303RT} -\frac{\Delta S}{2.303R}\right )
\end{equation}
where $\mathit{y}$ is the reciprocal of the coefficient of oxygen in the balanced redox reaction (Table \ref{tab:reac}), $\Delta H$, $\Delta S$, and $\Delta V$ are the change in heat of formation, entropy, and volume from reactants to products, $P_0$ is the reference pressure and $R$ is the gas constant (Table \ref{tab:param}).\\
\indent The oxidation-reduction reaction (Figure \ref{fig:Fug_final}) between siderite and diamond + w\"{u}stite equilibrates at a greater \fug than that of the iron to w\"{u}stite reaction  for the entire pressure and temperature range relevant from Mars to super-Earth mass mantles (0.1 - 250 GPa, 2000 - 5000 K) assuming an adiabatic geotherm characteristic of a silicate mantle \citep[Figure \ref{fig:Fug_final} inset, ][]{Dzie81}. This means that as the proportion of O increases relative to the rock-building cations (Fe, Mg, Si, C), iron will oxidize before carbon over the entire pressure and temperature range characteristic of Mars to super-Earth-sized planets.
\begin{deluxetable}{lccccc} 
\tabletypesize{\scriptsize}
\tablecolumns{6}
\tablecaption{Thermochemical parameters adopted in the fugacity model}
\tablehead{\colhead{Phase}&\colhead{$V_{0}$}&\colhead{$K_{0}$}&\colhead{$dK_{0}/dP$}&\colhead{$\Delta H^{0}_{f}$}&\colhead{$S^{0}$}\nl\colhead{}&\colhead{(cc mol$^{-1}$)}&\colhead{(GPa)}&\colhead{}&\colhead{(kJ mol$^{-1}$)}&\colhead{(J mol$^{-1} $K$^{-1}$)}}
\startdata
Diamond$^{1}$&3.42&442&4.0&1.897&2.38\nl
Enstatite$^{2,3}$&31.3&111.1&6.6&-1545.9&67.9\nl
Fayalite$^{3,4,5}$&46.3&123.9&6.0&-1479.4&151.0\nl
Iron-fcc$^{6,8}$&6.78&133.0&5.0&7.84&35.8\nl
Iron-hcp$^{7,8}$&6.73&165.0&5.33&7.7&34.4\nl
Magnesite$^{3,9,10}$&28.1&97.1&5.44&-1111.7&65.1\nl
Siderite-hs$^{11,13}$&29.4&117.0&4.0&-755&95.5\nl
Siderite-ls$^{12,13,*}$&25.59&146.7&4.0&-755&95.5\nl
Stishovite$^{3,14}$&14.0&312.9&4.8&-910.7&41.5\nl
W\"{u}stite$^{15,16}$&12.3&149.4&3.6&-267.3&57.59\nl
W\"{u}stite-8fold$^{16}$&11.9&137.8&4.0&-267.3&57.59\nl
Magnetite$^{17}$&44.56&217.0&4.1&-1115.73&146.15\nl
Periclase$^{18}$&11.24&156.0&4.35&-601.5&26.9
\enddata
\label{tab:param}
\tablerefs{\scriptsize 1. \citet{Mcsk72} 2. \citet{Ange94} \\3. \citet{Berm88} 4. \citet{Fuji81} 5. \citet{Robi78}  \\6. \citet{Camp09} 7. \citet{Mao90} 8. \citet{NIST} \\9. \citet{Lita08} 10. \citet{Robi82} 11. \citet{Lavi09} \\ 12. \citet{Matt07} 13. \citet{Mata00}  14. \citet{Pane03} \\15. \citet{McCam84} 16. \citet{Fisch11} \\17. \citet{Haav00} 18. \citet{Stix11} \\ *Pressure of spin transition: 45 GPa}
\end{deluxetable}
\begin{figure}
\plotone{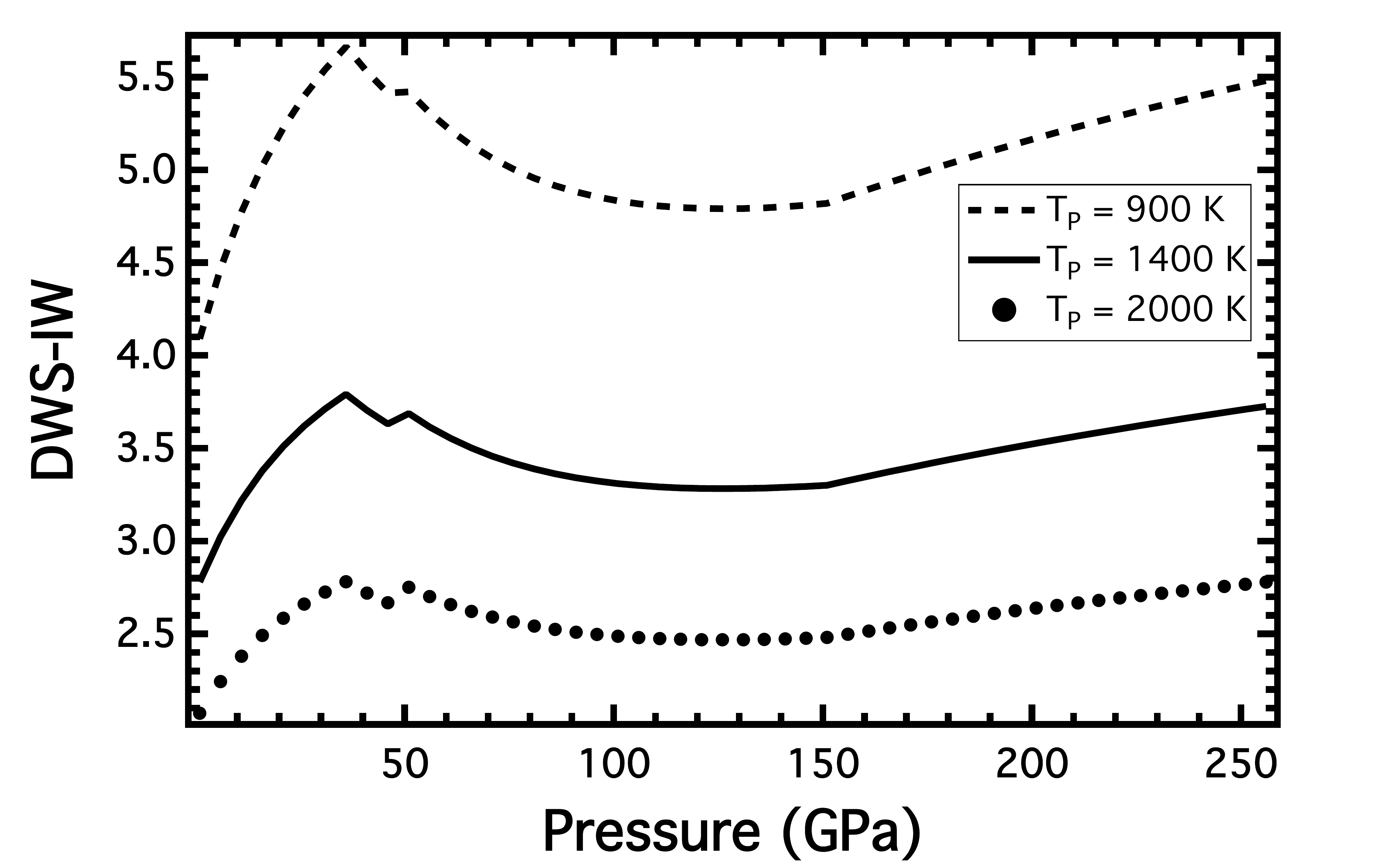}
\caption{Difference between CCO (DWS) and IW fugacities for three geotherms characteristic of a silicate mantle with potential temperatures of 900 (dashed), 1400 (solid) and 2000 (dotted) K. Potential temperature represents the temperature of a material when adiabatically brought to atmospheric pressure, thus negating any pressure induced temperature change. Potential temperatures provide us with a method with which to compare geotherms over very different pressure regimes. Furthermore, potential temperature allows us to compare planets with varying levels of heat producing radioactive elements (U, Th, K) or those of varying age with more of these elements decayed into their stable form.}
\label{fig:geo_diff}
\end{figure}
\\ \indent While the difference in \fug between the reactions decreases with increasing pressure, diamond oxidation to siderite never occurs at lower \fug than iron to w\"{u}stite for typical planetary mantle conditions, with the minimum separation of 2.54 log units at 150 GPa, and diverging at higher pressures. This phase stability divergence in \fug at pressures >170 GPa implies reduced mantles of carbon-bearing planets even considering variable Fe and FeO activities for Mg- and Si-bearing mantles. Furthermore, a change in our choice of mantle potential temperature ($T_P$) does not change the relative positions of these fugacity calculations (Figure \ref{fig:geo_diff}). \\
\indent Mg substitutes for Fe$^{2+}$ in many mineral systems. This Mg substitution into w\"{u}stite as periclase (MgO) or into siderite as magnesite (MgCO$_{3}$) shift both phase equilibria from the pure phase oxidation potential. To estimate these deviations away from the pure phase oxidation potential, we calculated the change in fugacity ($\Delta$\fugc) with respect to each oxidation potential by incorporating (1-$X$) mol Mg into the crystal structure where $X$ is the mol fraction of Fe in the system. Assuming this mixing is ideal and therefore follows Raoult's law, this correction is calculated by: 
\begin{align}
\Delta fO_2 &= fO_2^{mix}-fO_2^{pure} \nonumber\\
&= [X*fO_2^{DWS/IW} + (1-X)*fO_2^{MDP/PMg}]-fO_2^{DWS/IW} \nonumber\\
&= y\left [ (X-1) \left ( \frac{\Delta H+\int_{P_0}^{P}\Delta V dP}{2.303RT}-\frac{\Delta S}{2.303R} \right )_{DWS/IW}  \right. \nonumber\\
&+ \left. (1-X) \left ( \frac{\Delta H+\int_{P_0}^{P}\Delta V dP}{2.303RT}-\frac{\Delta S}{2.303R} \right )_{DPM/MP} \right. \nonumber\\
&+ \left. \frac{\Delta H_{mix}}{2.303RT}+\frac{\Delta S_{mix}}{2.303R}\right ]
\end{align}
where DPM and MP represent the magnesium variant of the DWS and IW equilibria respectively (Table \ref{tab:reac}). The entropy of mixing ($\Delta S_{mix}$) from Raoult's law is:\\
\begin{align}
\Delta S_{mix}=-R\left (X\ln X + (1-X)\ln (1-X)\right )
\end{align}
\noindent with $\Delta H_{mix}=0$ for an ideal solution. Assuming an Earth-like ratio of Mg/(Mg+Fe) = 0.8 corresponds to K$_{D}^{Mg}$(FeCO$_3$/FeO)=1, fugacity corrections were calculated for K$_{D}^{Mg}$=1 [(Fe$_{0.2}$,Mg$_{0.8}$)CO$_3$, (Fe$_{0.2}$,Mg$_{0.8}$)O], K$_{D}^{Mg}$=0.1 [(Fe$_{0.92}$,Mg$_{0.08}$)CO$_3$, (Fe$_{0.2}$,Mg$_{0.8}$)O], and K$_{D}^{Mg}$=10 [(Fe$_{0.08}$,Mg$_{0.92}$)CO$_3$, (Fe$_{0.8}$,Mg$_{0.2}$)O]. While Mg incorporation lowers the oxygen fugacity of both the IW and DWS reactions (Figure \ref{fig:mixing}), it does not cause the two oxidation potentials to cross suggesting iron will still oxidize before C over all pressures and temperatures explored here. \\
\section{Experimental Data}
\label{sec:Exp}
\indent The relative oxidation potential of the CCO and IW reactions in an iron-rich system are confirmed through high-pressure, high-temperature reaction experiments between equal proportions of metallic iron (Fe, Alfa Aesar, 1-3 $\mu$m, 98\%+ purity), w\"{u}stite (FeO, Alfa Aesar 99.5\%), and natural siderite ([Fe$_{0.74}$Mg$_{0.15}$Mn$_{0.08}$Ca$_{0.03}$]CO$_{3}$). Samples were loaded while in a nitrogen environment (O$_{2}$ < 0.1\%) into a laser-heated diamond anvil cell with a rhenium gasket and compressed to between 21 and 63 GPa and heated to 2150 (150) K for a minimum of 30 minutes. One sample compressed to 43 GPa was loaded using MgO as a pressure transmitting medium; whereas samples compressed to 21 GPa, 41 GPa, and 62 GPa were loaded using argon to serve as an inert pressure medium and insulator.  The experimental pressure range encompasses the spin transition of the ferric iron components \citep{Matt07}, and represents the conditions under which the oxidation potentials of the reactions are the most divergent (Figure \ref{fig:geo_diff}).   \\
\indent Resulting phase relations were determined via a combination of X-ray diffraction, Raman spectroscopy, and Energy Dispersive X-ray (EDX) spectroscopy of extracted foils measuring 10x20x0.1 $\mu$m using focused-ion beam milling (FIB).  These techniques provide bulk (X-ray diffraction) and 5 $\mu$m-scale resolution (Raman) mineral phase identification. X-ray diffraction, while averaging over the full sample, also measures the lattice volumes of each phase providing a measurement of oxygen vacancies in the w\"{u}stite. The STEM EDS measures provide grain-by-grain mineral associations and compositions on a 5 nm length scale. \\
\begin{figure*}
  \begin{center}
    \leavevmode
      \plotone{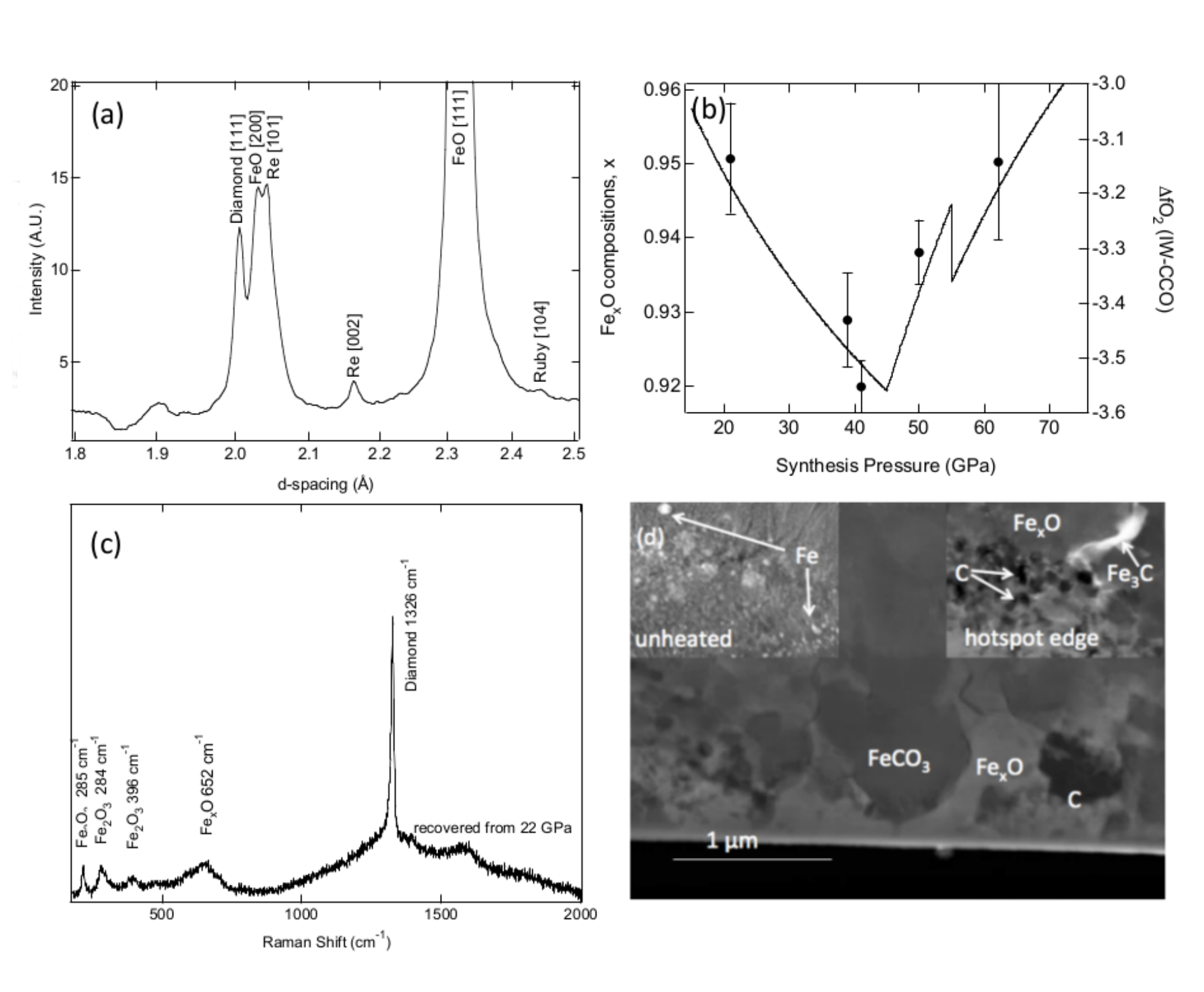}
       \caption{Reactions between FeO, FeCO$_3$, and Fe at 21-63 GPa and 2150(150) K show the oxidation of iron and reduction of carbon. (a) X-ray diffraction at 43 GPa shows 
		the presence of diamond and w\"{u}stite with no evident siderite.  (b) The w\"{u}stite formed is deficient in iron as measured by the unit-cell volume of the recovered sample, where the relationship between the Fe/O ratio $x$ varies along with the differences in the IW and CCO fugacities.  (c) Raman spectroscopy of a sample recovered from 22 GPa (outside the diamond cell) confirms the formation of diamond, along with residual siderite from unheated portions of the sample volume, where the presence of Fe$_2$O$_3$ is interpreted as a consequence of surface oxidation in the recovered sample. (d) TEM image of a foil extracted by focused-ion beam milling from a sample recovered from 63 GPa and 2200 K with 10-200 nm diamond grains surrounded by FeO and associated with carbonate, whereas unheated portions of the sample show iron grains (inset, top left), and portions of the sample near the edges of the hotspot ($\sim$1500 K, inset top right) show smaller diamond grains and iron-carbide with w\"{u}stite.}
	\label{fig:Exp_fig}
  \end{center}
\end{figure*}
Each high-pressure experiment in the Fe-FeO-\sid  system results in diamond formation via carbon reduction and iron oxidation as marked by metallic iron consumption (Figure \ref{fig:Exp_fig}). The equilibration results place the oxidation potential of this system above the IW stability and below the DWS stability. The unit-cell volumes of the w\"{u}stite, Fe$_{x}$O, indicate variable iron content  with $x$ between 0.91 and 0.96 \citep{McCam84}. Iron deficient w\"{u}stite is charge balanced through the oxidation of iron to Fe$^{3+}$ as defects, where the greatest oxidation is observed at 41 GPa where the difference between the IW and DWS is near a maximum (Figures \ref{fig:geo_diff}, \ref{fig:Exp_fig} top right). We interpret this as a function of the difference between the \fug of the system and the iron-w\"{u}stite oxidation potential. A few 50-100 nm grains of iron carbide (consistent with Fe$_{3}$C) observed in TEM and X-ray diffraction were found in the 63 GPa sample near the laser-heating edges also associated with FeO.  In this case, regions of the sample not as hot were more reducing, thereby placing the system closer to the IW reaction stability, leading to the reaction of the diamond and iron to iron carbides.  These experimental results confirm the modeling from thermodynamic parameters, effectively capturing details such as the iron spin transitions in the iron carbonate and oxide.
\section{Planetary Mantle Model}
\label{sec:model}
\begin{figure*}
  \begin{center}
    \leavevmode
      \epsfxsize=15cm\plotone{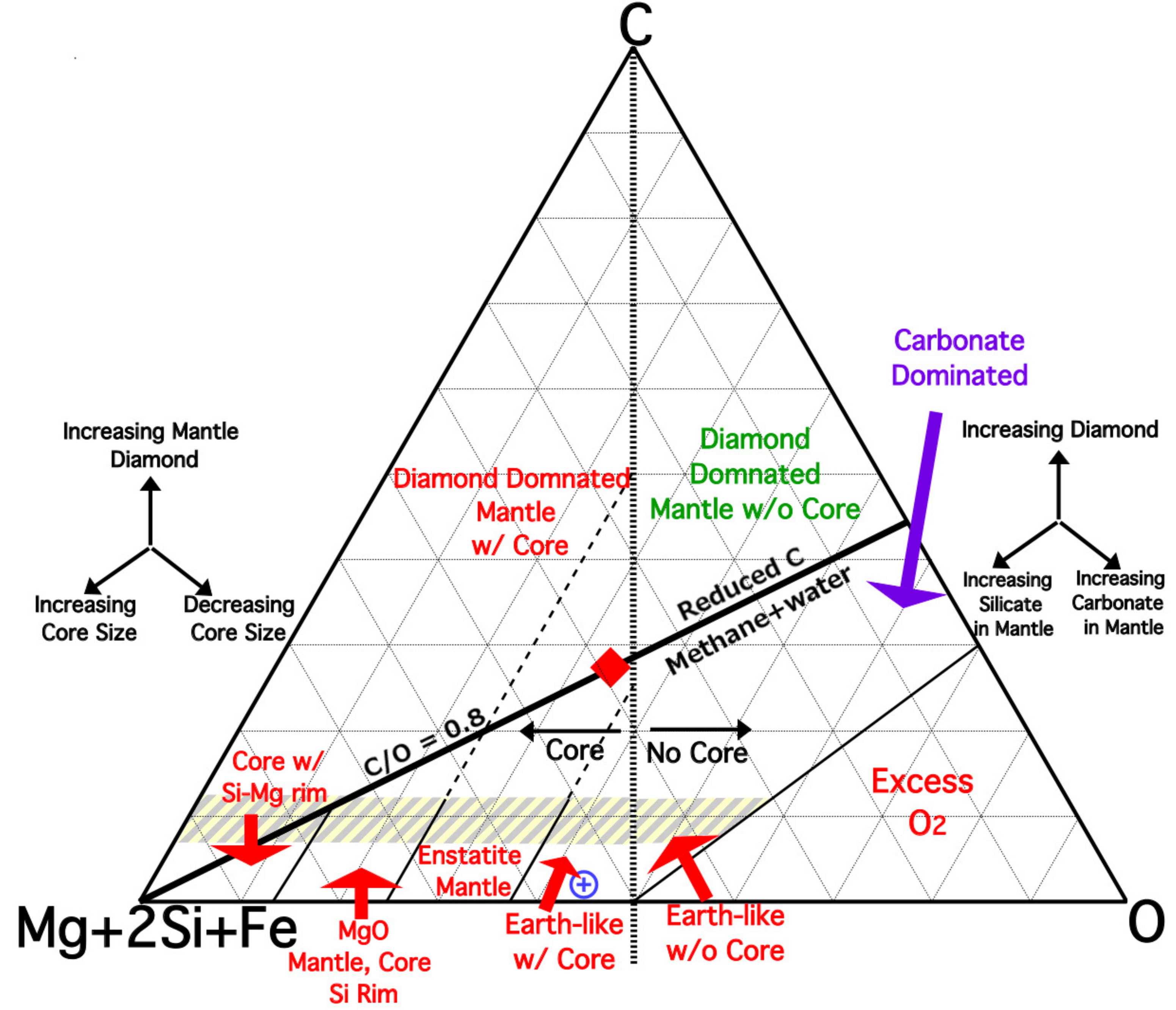}
	\caption{Ternary diagram for the C-(Mg+2Si+Fe)-O system. The Earth is shown as a blue cross and HD19994 as a red diamond. The exact position of the diamond/no diamond line on the core-free side of the ternary depends on the specific Fe/(Mg+2Si+Fe) ratio, in which planets with more Fe relative to the other cations able to stabilize more \sidc, and thus reduce the amount of diamond present in the mantle. The shaded region represents the maximum C concentration within a planet so as to not affect mantle dynamics and subsequent habitability. This region can move up or down with higher or lower mantle potential temperature.}
	\label{fig:Ternary}
	\end{center}
\end{figure*}

\begin{deluxetable}{lc|cc|ccc}
\tablecolumns{7}
\setlength{\tabcolsep}{0.00002in} 
\tabletypesize{\scriptsize}
\tablecaption{Distribution of minerals of modeled carbon-enriched planet orbiting HD19994 \citep{Bond10} as calculated by our simple mantle model. Included is a simplified Earth  mineral composition with bulk element data from \citet{McD03} normalized to a composition that does not include other major cations such as Al and Ca.}
\tablewidth{8cm}
\tablehead{\colhead{}&\colhead{\textbf{Earth}}\vline&\multicolumn{5}{c}{\textbf{HD19994}}\nl\hline 
\colhead{}&\colhead{Bulk}\vline&\colhead{Mantle}&\colhead{Core}&\colhead{Mantle}&\colhead{}&\colhead{}\nl 
\colhead{Mineral}&\colhead{mol$\%$}\vline&\colhead{mol$\%$}&\colhead{mol$\%$}&\colhead{Phase}&\colhead{mol$\%$}&\colhead{vol$\%$}}
\startdata
Fe &27.8\footnotemark& - &7.4&(Mg,Fe)SiO$_{3}$&25.9&63.3\nl
FeC&-&-&1.9*10$^{-6}$&(Mg,Fe)O&11.8&12.4\nl
Fe$_{3}$C&-&-&5.6*10$^{-6}$&C&62.3&24.3\nl
Fe$_{7}$C$_{3}$&-&-&0.1&&&\nl
SiO$_{2}$&32.1&19.1&-&&&\nl
MgO&35.5&19.6&-&&&\nl
FeO&4.26&8.1&-&&&\nl 
C&0.34&45.7&-&&& 
\enddata
\footnotetext[1]{Contained in the core}
\label{tab:comp} 
\end{deluxetable}
\indent As a planet cools from an initial magma ocean, the total oxygen budget relative to rock-building cations determines the final oxidized mineral assemblages present. As shown in our thermodynamic calculations and experiments, the CCO (DWS) reaction is at a greater oxidation potential than the IW reaction over all pressures and temperatures characteristic of a Mars to super-Earth-sized terrestrial planet. This means that at a given depth and temperature, Mg will oxidize first, followed Si, Fe, and finally C. These simple oxide reactions will cease when all oxygen is consumed, with more complex mineral assemblages such as Mg-Fe perovskite forming from the oxides. With planets that form hot, the high-density, relatively low melting temperature materials, i.e. iron and iron carbide, will gravitationally segregate into a planetary core.\\
\indent With a lack of oxygen to fully oxidize all of the cations, planets having (Mg+2Si+Fe+2C)/O > 1, will have an excess of a reduced species. We define planets with these compositions as those containing central cores (Figure \ref{fig:Ternary}). For example, the chondridic Earth has (Mg+2Si+Fe+2C)/O = 1.27 \citep{McD95}. Assuming stoichiometric oxidation first of Mg, then Si, and finally Fe, 96\% of the total oxygen budget is consumed by Mg and Si. The remaining 4 atom\% oxidizes only 13 atom\% of the total iron leaving an excess of $\sim$5 mol\% reduced iron. If all of this metallic iron is assumed to segregate into the Earth's core, it accounts for $\sim$87\% of the core's true mass with the remaining 13\% due to nickel and the incorporation of minor elements into the core \citep{McD03}. With the oxygen supply exhausted, carbon too will be present in its reduced form accounting for only 0.32 mol\% of the Earth. While an extremely oversimplified model of planetary differentiation, simple oxidation reactions in the C-(Mg+2Si+Fe+2C)-O system largely account for the gross structure and mineralogy of the Earth.To fully determine the mineralogy of the Earth from a simple model such as this, the proportions of the other minor rock building cations (Ca, Al) and the oxidation potentials of reactions containing these elements, both of which are outside of the scope of this paper.\\
\indent Planets with (Mg+2Si+Fe+2C)/O < 1, on the other hand, will have an excess of oxygen relative to the dominant cations (Figure \ref{fig:Ternary}). With no metallic Fe present, a core will not form on a planet of this composition \citep{ElkT08}. Furthermore, if these planets contain large amounts of O and C relative to (Mg+2Si+Fe), there will be a sufficient excess of C to retain both diamond and oxidized species in its mantle (Figure \ref{fig:Ternary}). \\
\indent The Earth represents a case of low C abundance relative to the major cations and oxygen, however recent N-body simulations have suggested the presence of planets with relatively high C abundances ($\sim$30 atom\%) in the C-(Mg+2Si+Fe)-O system \citep{Bond10}. One possible example of such high-C planets is that of those modeled to be orbiting HD19994 (Table \ref{tab:compcompare}). For example, in this planet's mantle all of the Si and Mg are oxidized to SiO$_2$ and MgO, accounting for 88\% of the planet's oxygen budget, leaving enough oxygen to oxidize 49\% of the planet's Fe into FeO. With the oxygen supply exhausted, the remaining carbon can take three forms: it can alloy itself with the excess iron to form a carbide or reduce to diamond. \\
\indent While the oxidation state of carbon depends on the underlying chemistry of the system, carbon's ability to alloy with iron at temperatures below their respective liquidi, is dependent on the ratio of free iron to reduced carbon \citep{Dasg10}. Adopting the high-pressure phase relationships in the Fe-C system at 50 GPa \citep{Lord09} and assuming fractional crystallization, we find that roughly 1.4 wt\% carbon alloys with iron forming the carbides Fe$_7$C$_3$, Fe$_3$C and FeC accounting for 0.1, 6$\times$10$^{-6}$, 2$\times$10$^{-6}$ mol\% respectively. While these values are based on data at pressures and temperatures indicative of the Earth, the solubility C into Fe is expected to increase as pressure and temperature increase, therefore these values should be considered lower limits. The densities of both iron and iron carbide are significantly greater than diamond and silicates and have lower melting temperatures. We assume then that accretionary melting of these planets will differentiate the planet into a iron/iron carbide core accounting for $\sim$13\% of the planet's mass surrounded by a mantle made of diamond (62 mol\%; 24 vol\%) and silicate minerals (Table \ref{tab:comp}). This model represents a maximum core size. While the Earth also has (Mg+2Si+Fe+2C)/O > 1, it creates far less diamond accounting for only 0.1 wt\% of the mantle \citep{McD03}. While this calculation is meant only as an extremely simplified example, it points to the importance of (Mg+2Si+Fe+2C)/O in a planet's overall bulk mineral composition. \\
\indent For stellar planet hosts with C/O > 0.8, the dominant accretionary form of C is in the form reduced carbides or graphite \citep{Lari75,Bond10}. These carbides have condensation temperatures between 1400 and 1600 K, thus providing a relatively large reservoir of reduced C to an accreting terrestrial planet. However, as C/O decreases below 0.8, C can condense in three forms, CO and CO$_2$ ices or graphite.  This condensation occurs along three end-member chemistries \citep{Lodd03}. Under equilibrium, CO is the dominant phase in the protoplanetary disk at high temperatures with methane replacing it as temperature decreases. At equilibrium, methane is produced via the reaction:
\begin{equation}\label{eq1}\ce{CO +3H_2 = CH_4 + H_2O}\end{equation}
at $\sim$650 K. At the pressures and temperatures in a disk, however, the reaction from carbon monoxide to methane occurs very slowly \citep{LewP80}. In kinetically inhibiting methane formation, two non-equilibrium chemistries determine the condensing phase of C:
\begin{equation}\label{eq2}\ce{CO = C(graphite) + CO_2}\end{equation}
with graphite condensation beginning at $\sim$626 K, well above the water and CO$_2$ ice condensation temperature and
\begin{equation}\label{eq3}\ce{2CO +H_2O = CO_2 + H_2}\end{equation}
 In the case of reaction \ref{eq3}, the vapor pressure of water controls the direction of the reaction. At temperatures greater than the condensation temperature of water, the reaction stabilizes CO$_2$ and H$_2$. As the reaction progresses, the water vapor pressure drops, thus lowering the condensation temperature of water to 121 K, still above that of CO$_2$ ice. As water begins to condense, the reaction begins to move towards the left, with the condensing water removing O from the system and causing the gas to become extremely saturated in C, particularly CO. Even if reaction \ref{eq1} is kinetically inhibited, near the condensation temperature of water, reaction \ref{eq3} will still produce graphite via reaction \ref{eq2} at  temperatures below 626 K. Graphite condensation will begin before CO$_2$ ice, such that only with significant radial mixing will a planet accrete both graphite and oxidized species. It holds then, that even systems with C/O < 0.8 can accrete reduced C, regardless of water content, with the specific (Mg+2Si+Fe+2C)/O depending on the exact composition of the planetary nebula. 
\section{Impact of Carbon on Interior Dynamics}
\begin{figure}
\includegraphics[width=9cm]{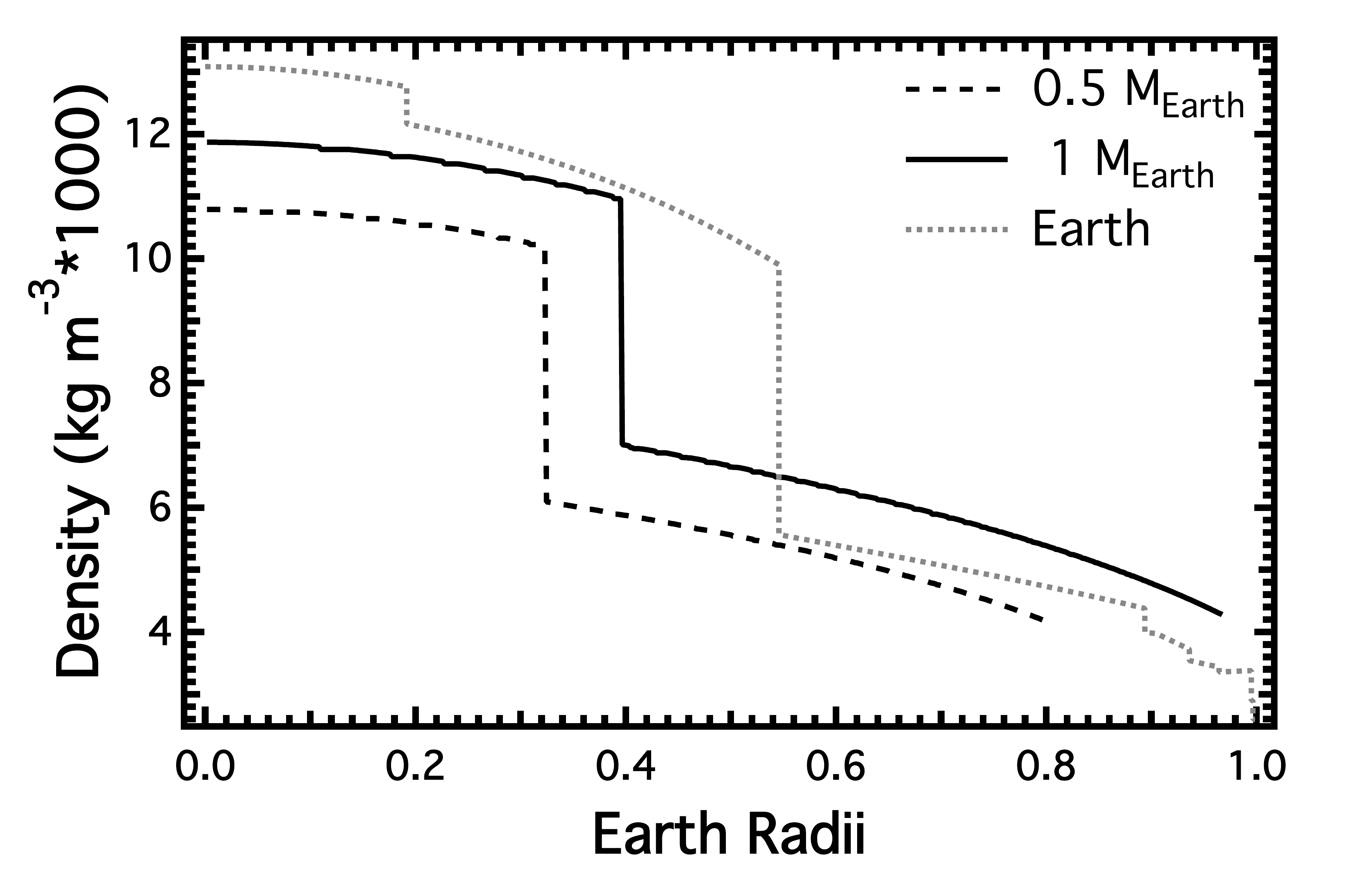}
\caption{Density-radius relations for 0.5 (dashed) and 1 (solid) Earth mass planets with an equal molar fraction of magnesium perovskite (Mg/Si = 1, no ferropericlase) and C in its mantle and 13 wt\% Fe in the core. Earth as calculated from the Preliminary Reference Earth Model \citep{Dzie81} is shown in gray dotted for reference.}
\label{fig:density}
\end{figure}

\begin{figure}
\includegraphics[width=9cm]{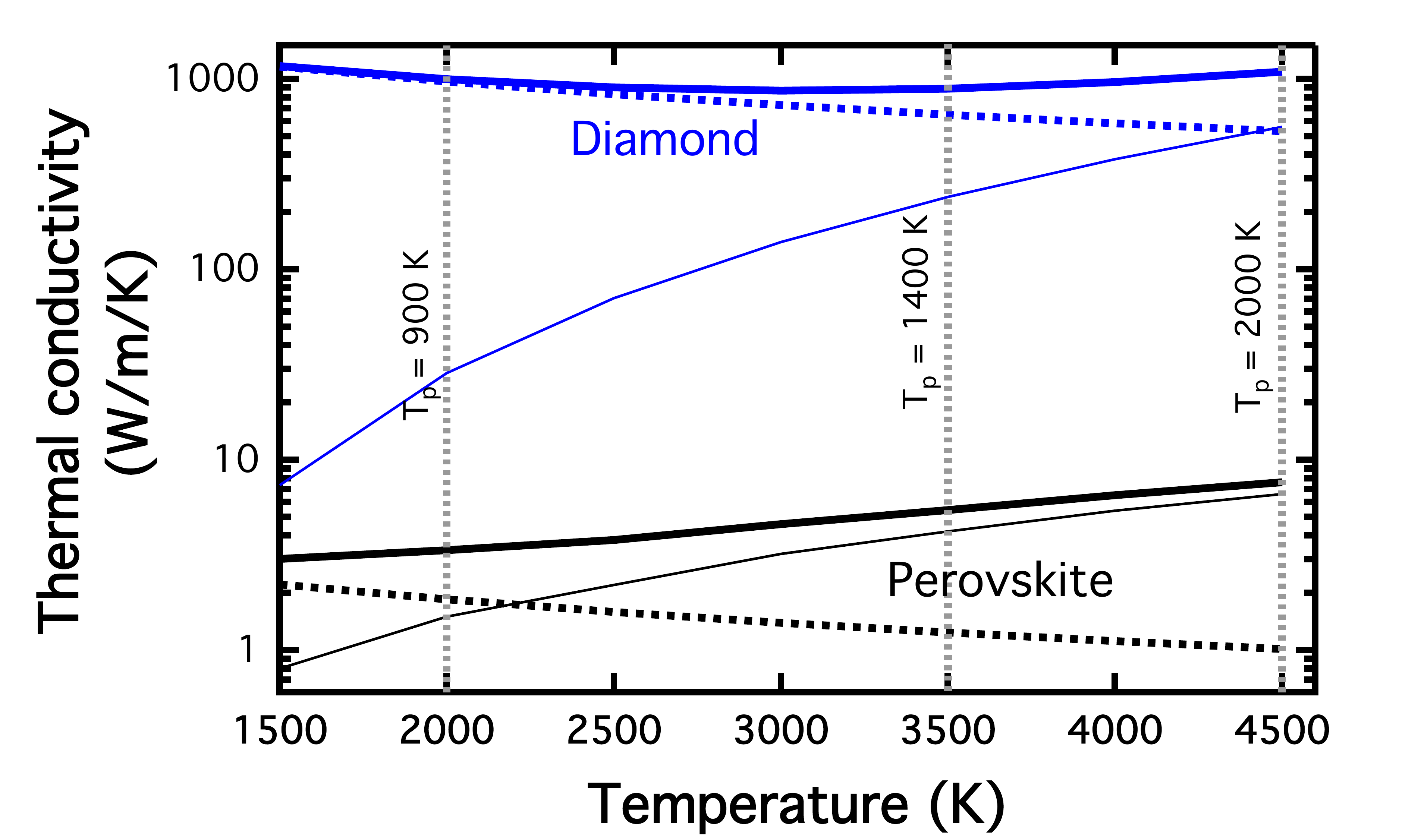}
\caption{Diamond (blue) and perovskite (black) lattice (dashed) and radiative (thin solid) thermal conductivity. Bold lines represent the sum of lattice and radiative thermal conductivity for each mineral. Choice of temperature to determine end-member values for our Rayleigh number calculation are shown as gray dashed lines.}
\label{fig:thermalcond}
\end{figure}

\begin{figure}
\includegraphics[width=9cm]{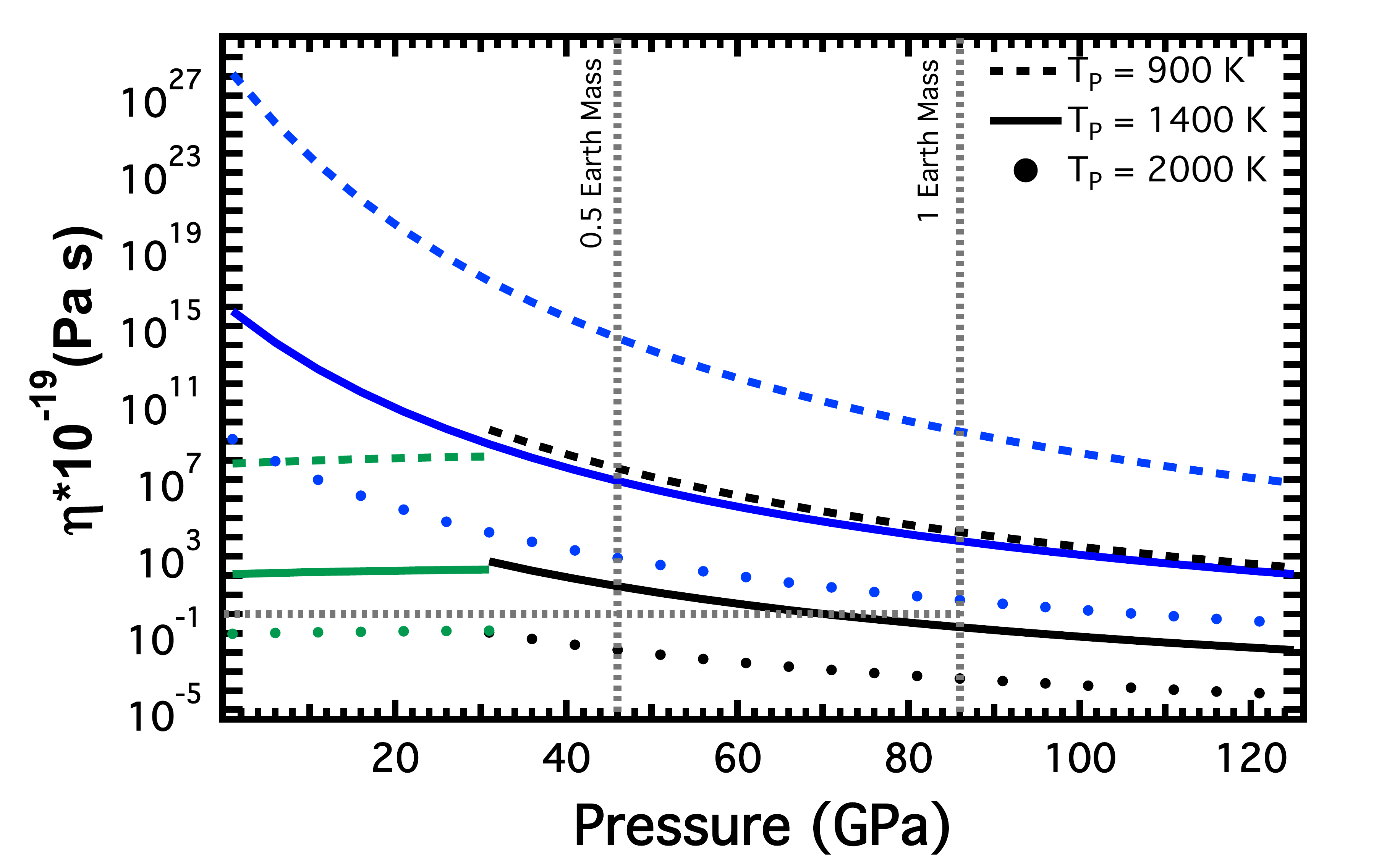}
\caption{Lattice diffusion creep viscosity profiles for diamond (blue), olivine (green) and perovskite (black) along an adiabatic Earth-like geotherm adopting a mantle potential temperatures of 900 (dotted), 1400 (solid) and 2000 (dashed) K. The reference average Earth viscosity chosen as 10$^{19}$ Pa s \citep[][gray horizontal line]{Kore11}. The vertical gray line represents the chosen pressure for our Rayleigh number calculations (Figure \ref{fig:RayleighNumber}). }
\label{fig:viscosity}
\end{figure}

\indent Planetary mantles containing a significant volume fraction of diamond will have dramatically different dynamics and thermal evolution than Earth-like, silicate-dominated planets. While plate tectonic-like regimes are a more complex question of fault strength, surface gravity, and the presence of liquid water \citep{Vale09, Vale07,Crow11,VanH11, Tack13}, the Rayleigh number provides, to first order, a measure of whether interior mantle convection will occur in these planets. There is no mechanism to induce or sustain plate tectonics without interior mantle convection. The Rayleigh number ($Ra$) of a system is: 
\begin{equation}
Ra=\frac{g\rho_{0}^2\alpha \Delta TD^{3}C_{p}}{k\eta _{0}}\
\end{equation}
\noindent{where $\mathit{g}$ is the acceleration due to gravity, $\mathit{\rho _{0}}$ is the density of the mantle, $\mathit{\alpha}$ is the thermal expansion of the materials present, $\mathit{\Delta T}$ is the change in temperature across a boundary layer, $\mathit{D}$ is the radius of the boundary layer, $\mathit{C _{p}}$ is the specific heat of the system, $\mathit{k}$ is the thermal conductivity and $\mathit{\eta _{0}}$ is an average viscosity. Above the critical Rayleigh number \citep[Ra$_{crit}$ = 10$^3$,][]{Schu79,Schu80}, mantle interior heat will be more efficiently transported via convection, with conduction being the dominant form of heat transport for those planets with Rayleigh numbers below the critical value. }  \\
\indent To determine the Rayleigh number for planets of variable C content, we calculate the values of $\mathit{g}$, $\mathit{\rho_0}$ and $\mathit{D}$ through the density-radius relationship using the coupled differential equations: 
\linebreak
\noindent{the mass within a sphere, }
\begin{equation}
\label{MR1}
\frac{dm(r)}{dr}=4\pi r^{2}\rho(r)
\end{equation}
\noindent{the equation of hydrostatic equilibrium}
\begin{equation}
\label{MR2}
\frac{dP(r)}{dr}=\frac{-Gm(r)\rho(r)}{r^2}
\end{equation}
and the isothermal, Birch-Murnaghan equation of state:
\begin{equation}
\label{MR3}
P(r) = f(r)
\end{equation}
where $G$ is the gravitational constant, $\rho$ is the density, $P(r)$ is the pressure at radius $r$ and $m(r)$ is the mass within the sphere of radius $r$ (Figure \ref{fig:density}). \\
\indent We adopt Earth-like values of $\mathit{C_p}$ and $\mathit{\alpha}$. In the high temperature limit, $\mathit{C_p}$ is a constant, and $\mathit{\alpha}$ will vary by less than an order of magnitude due to bulk mineralogical differences, thus having very little effect on the overall Rayleigh number. \\
\indent Diamond has a lattice thermal conductivity two orders of magnitude larger than that of a bulk silicate similar to Earth \citep[e.g.][]{Osak91,Pane01}.  Highly transparent to visible and near infrared light, diamond's radiative thermal conductivity is three orders of magnitude higher than silicates \citep[Figure \ref{fig:thermalcond},][]{Kepp08}. In modeling thermal conductivity we adopt published values for the radiative transfer of perovskite at 125 GPa \citep{Kepp08}, lattice conductivity \citep{Osak91}, and temperature dependence \citep{Pane01}. Thermal conductivity of diamond is assumed to be 2200 W m$^{-1}$ K$^{-1}$ at 300 K, with the same temperature dependence as perovskite.  Radiative transport was then calculated as in \citet{Kepp08}, with an absorption spectrum as measured on a gem quality diamond, thereby representing an upper bound for radiative thermal conductivity.\\
\indent Adopting the power-law creep formalism of \citet{Kara93}, we assume that perovskite viscosity scales relative to dry olivine diffusion creep using the equation for strain rate, $\dot{\varepsilon}=A(\sigma/\mu)(b/d)^{m}\exp[-(E^{*}+PV^{*}/RT)]$, where $\mathit{A}$ is the preexoponential factor, $\mathit{\sigma}$ is the shear stress, $\mathit{\mu}$ is the shear modulus, $\mathit{b}$ is the length of the Burgers vector, $\mathit{d}$ is the grain size, $\mathit{m}$ is the grain size exponent, $\mathit{E^{*}}$ is the activation energy, $\mathit{V^{*}}$ is the activation volume, $\mathit{R}$ is the gas constant, $\mathit{P}$ is the pressure and $\mathit{T}$ is the temperature along adiabatic geotherms as described by the surface potential temperature. For a silicate mantle similar to the Earth, this is $\sim$1400 K, which we adopt as our intermediate geotherm, with 900 and 2000 K representing cold and hot examples, respectively.\\
\indent The effective viscosity is then $\sigma/\dot{\varepsilon}$ yielding $\eta=(\mu/A)(d/b)^{m}\exp[(E^{*}+PV^{*}/RT)]$. Viscosity model parameters are as in Table \ref{tab:visc}. \\
\indent For our calculations, we adopt as end-member average viscosities the value at pressure at one-half the total planetary radius. This corresponds to roughly 46 GPa and 86 GPa for a half and one Earth mass planet respectively. For an Earth-mass planet (86 GPa), we find diamond viscosity to be 5 orders of magnitude higher than an average viscosity for the Earth \citep[$\eta _0 \sim 10^{19}$ Pa s, ][]{Kore11} for each of our geotherms (Figure \ref{fig:viscosity}). In 1/2 Earth-mass case (46 GPa), this difference increases to 5-7 orders of magnitude higher. For the mantle silicate species, these calculated viscosities are within an order of magnitude of those of \cite{Tack13} and \cite{Amma10} over the pressure and temperature range explored here. Composite viscosities and thermal conductivities of a diamond-perovskite mixture are calculated from the end-member values using Voight-Reuss-Hill averaging \citep{Watt88}. \\
\begin{deluxetable}{lccc} 
\tabletypesize{\scriptsize}
\setlength{\tabcolsep}{.001cm} 
\tablecolumns{4}
\tablecaption{Viscosity parameters adopted for our model}
\tablehead{&\colhead{Diamond}&\colhead{Perovskite}&\colhead{Reference}}
\startdata
A(s$^{-1}$)&1.2*10$^{16}$&2.67*10$^{17}$&Scaled to diffusivities\\
m&2.5&2.5&Assumed same as olivine\\ 
E$^{*}$(kJmol$^{-1}$)&655&501&1,2\\
V$^{*}$(cm$^{3}$mol$^{-1}$)&1.5&2.1&3\\
d(m)&0.001&0.001&Scaled to xenolithgrains\\
b(m)&1.75*10$^{-10}$&6*10$^{-10}$&scaled to olivinelattice\\
$\mu$(Pa)&1.36*10$^{11}$&1.84*10$^{11}$&4,5
\enddata
\tablerefs{\scriptsize 1. \citet{Koga05} 2. \citet{Dobs08} 3. \citet{Holz05} 4. \citet{Kepp08} 5. \citet{Yega89} }
\label{tab:visc}
\end{deluxetable}

\begin{figure*}
  \begin{center}
    \leavevmode
      \epsfxsize=15cm\plotone{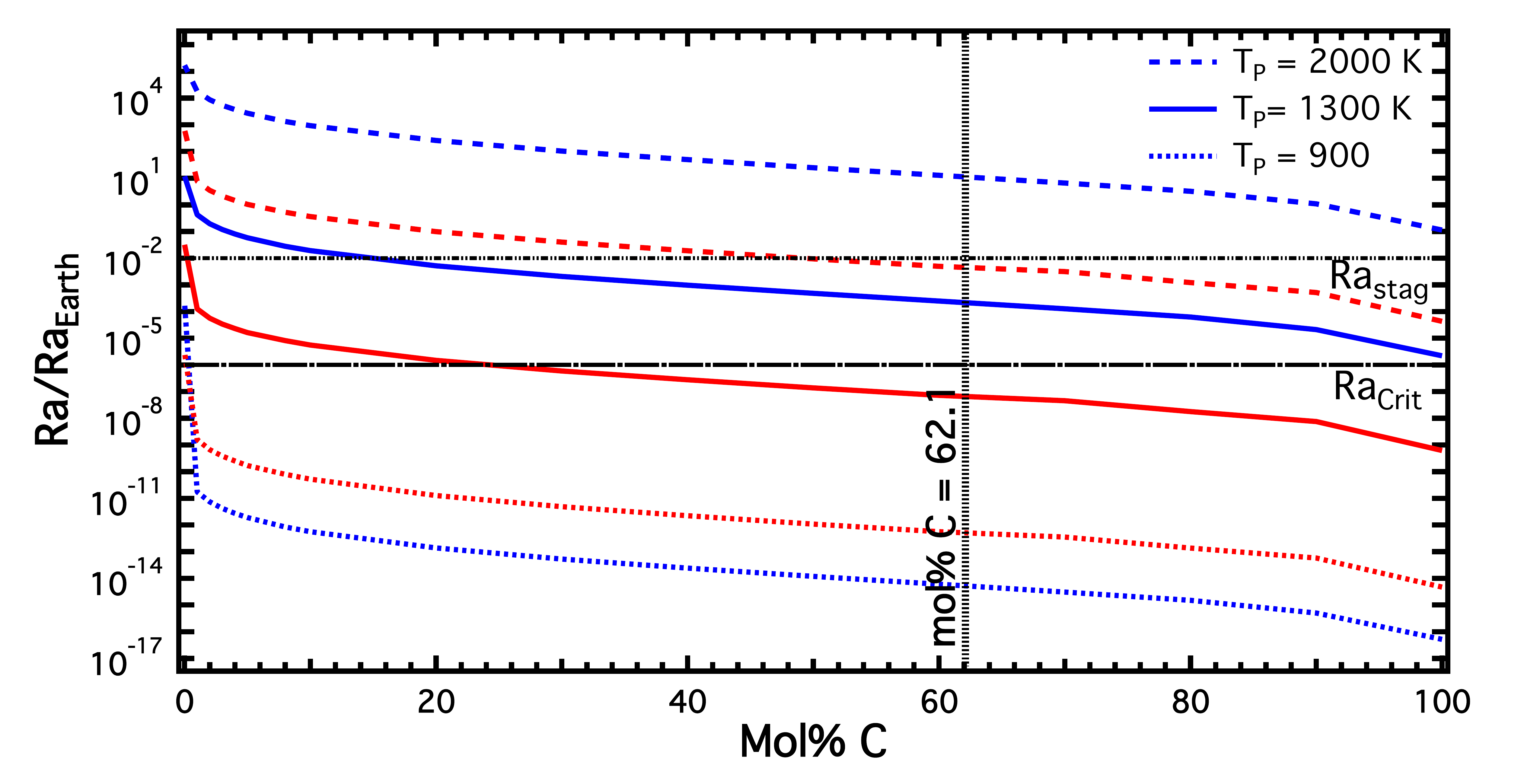}
	\caption{Calculated Rayleigh versus volume percentage of diamond relative to Earth's Rayleigh number ($\sim$10$^{9}$) at 46 GPa for 0.5 (red), and 86 GPa for 1 (blue) Earth mass planets with at three different mantle potential temperatures: 900 (dotted), 1400 (dash-dotted) and 2000 K (dashed). The critical Rayleigh number is shown in black (long dashed) as well as the mol\% of the planet considered in Table \ref{tab:comp} (thin dotted).}
	\label{fig:RayleighNumber}
	\end{center}
\end{figure*}
\indent For planets with internal heat production comparable to or greater than the Earth ($T_p \geq$1400 K), the calculated Rayleigh numbers are above the critical values   while those with lower heat productions fall below the critical Rayleigh number regardless of C content or planet size (Figure \ref{fig:RayleighNumber}). While a planet may be above this critical Rayleigh number, the convective vigor must be greater than  $\sim$10$^7$  to support plate tectonics, below which it will be in the stagnant lid regime \citep{Kore11}. The proportion of diamond present in the Earth's mantle is too small to have any noticeable effect on bulk mantle dynamics. Therefore we define the difference between an Earth-like and high-carbon planet as one in which the excess C reduces the Rayleigh number below the stagnant lid value of $\sim$10$^7$ and therefore begins to affect bulk planetary dynamics. This occurs for an Earth-mass planet along an intermediate geotherm at $\sim$10 mol\% or 3 atom\%. Planets above this C concentration will have very different mineralogies than Earth and consequently occupy a wider range of dynamical regimes. \\

\section{Conclusion}
With no more than short-lived surface-to-interior mass flux in a planet with C > $\sim$3 atom\%, plate tectonics and any related mechanisms for cycling deep carbon or water into and out of the mantles will be limited. Furthermore, a stagnant lid or lack of interior dynamics will disable surface volcanism without significant tidal heating, thus hindering the development and preservation of an atmosphere, vastly lowering its habitability. Extending the mineralogical model to the surface, these planetary crusts will contain mostly graphite and mafic silicates, creating a low surface albedo, absorbing the majority of the energy of the parent star, leading to relatively high surface temperatures for a given distance from the star. With no transport of volatiles into the planetary mantle combined with high surface temperatures and little to no atmosphere present, creating and retaining surface oceans will be nearly impossible, thus producing planets inhospitable to life as we know it. \\

\acknowledgments
This work is supported by NSF CAREER (EAR-60023026) and PRF (47664-G8). X-ray and infrared experiments were performed at beamlines X17-C and U2A at the National Synchrotron Light Source at Brookhaven National Lab and supported by COMPRES. EDX and TEM analysis were performed at the Campus Electron Optics Facility at Ohio State University, and Raman spectroscopy at the Analytical Spectroscopy Laboratory in the Department of Chemistry at Ohio State University. The authors would like to thank Jingzhu Hu for her help with the X-ray diffraction studies. 
\bibliographystyle{apj}
\bibliography{Diamond_ApJ}
\end{document}